\documentclass[reprint, aps,prd,twocolumn,showpacs,showkeys, 10pt, longbibliography, nolinenumbers, floatfix]{revtex4-2}

\usepackage{amsmath}
\usepackage{graphicx}
\usepackage{float}
\usepackage{dcolumn}
\usepackage{multirow}
\usepackage{natbib}
\usepackage{changes}
\usepackage[colorlinks = true,linkcolor = blue,urlcolor  = blue,citecolor = blue,anchorcolor = blue]{hyperref}
\usepackage{enumerate}
\usepackage{verbatim}
\usepackage{orcidlink}

\begin{document}

\title{Neutron Dark Decay and Exotic   Compact Objects}

\author{M. Vikiaris$^1$}
\author{V. Petousis$^2$}
\author{M. Veselsky$^2$}

\author{Ch.C. Moustakidis$^1$}


\affiliation{$^1$Department of Theoretical Physics, Aristotle University of Thessaloniki, 54124 Thessaloniki, Greece
\\
$^2$Institute of Experimental and Applied Physics, Czech Technical University, Prague, 110 00, Czechia}

\begin{abstract}

 Recent measurements of the compact star XTE J1814-338, with a mass of
 $M=1.2_{-0.05}^{+0.05}\ M_{\odot}$
and a radius of $R=7_{-0.4}^{+0.4} \ {\rm Km}$
alongside those of HESS J1731-347, which has a mass of
 $M=0.77_{-0.17}^{+0.20}\ M_{\odot}$ and a radius of $R=10.4_{-0.78}^{+0.86} \ {\rm Km}$, provide compelling evidence for the
potential existence of exotic matter in neutron star cores. These observations offer important
insights into the equation of state of dense nuclear matter. Recently,  Fornal and  Grinstein, in order to overcame  the discrepancy between the neutron lifetime
measured in beam and bottle experiments, proposed the existence of neutron dark  decay.  In the present work, an effort is made to connect the interpretation of the above exotic compact objects with the possible existence of dark particles, assumed to be products of neutron dark decay.
Our hypothesis offers an advantage over comparable proposals, as the coexistence of dark matter and hadronic matter within neutron stars emerges from an intrinsic mechanism, thereby obviating the need to invoke external merger-related processes.
It is still unclear to what extent the proposed dark decay of the neutron is affected by the extreme environment within neutron stars. Within this framework, we examined the case in which a mechanism suppressing  the dark neutron decay becomes operative at densities few times above nuclear saturation density. We found that the  proposed  alternative explanation accommodates the simultaneous existence of neutron dark decay while consistently predicting both the two solar mass limit and the presence of compact objects with subsolar masses.


\end{abstract}

\maketitle

\section{Introduction}

One of the most long-standing problems in nuclear
physics is the discrepancy between the neutron lifetime
measured in beam and bottle experiments.  In the bottle method,
neutrons are trapped, making it sensitive to the total neutron width, yielding a slightly
shorter lifetime of $\tau_n^{\rm bottle}=
879.6\pm 0.6$ s \cite{Pichlmaier-2010,Serebrov-2005,Steyerl-2012}. In contrast, the beam method measures
the number of protons produced through $\beta$-decay in a fixed volume from a neutron beam,
resulting in a longer lifetime of $\tau_n^{\rm beam}=
888.0\pm 2.0$ s \cite{Arzumanov-2015,Byrne-1990,Yue-2013}. This discrepancy  led  Fornal and  Grinstein to propose the existence of neutron dark  decay~\cite{Fornal-2018}. In particular, they proposed as an explanation for this anomaly, a dark decay channel for the neutron, involving one or more dark matter particles in the final state (for more details see also Ref.~\cite{Fornal-2023}).

Following the hypothesis of   Fornal and  Grinstein~\cite{Fornal-2018},  a series of papers studied the effect of the   neutron dark  decay on the properties of neutron stars (NSs) and finite nuclei~\cite{Baym-2018,Gil-2024,Motta-2018a,Motta-2018b,Husain-2022a,Grinstein-2019,McKeen-2018,Husain-2023,Husain-2025,Shirke-2023,Shirke-2024,Das-Burgio-2025,Tan-2019,Cline-2018,Fornal-2020,Darini-2023,Strumia-2022,Ejiri-2019,Ivanov-2019,Veselsky-2025,Harris-2025,Divaris-2025,Vikiaris-2024,Vikiaris-2025} (see also the review papers~\cite{Zhou-2023,Gardner-2023,Bramante-2024,Grippa-2025}). A common conclusion of the above studies is that dark matter (DM) must exhibit sufficiently repulsive interactions to overcome the softening of the equation of state caused by the addition of  a
second species, thus accommodating $2 M_{\odot}$ neutron stars.
According to Ref.~\cite{Baym-2018} such a scenario makes very specific demands on the dark
matter self-interaction strength as a function of the dark
matter fermion density. Moreover, in Ref.~\cite{Grinstein-2019} the authors demonstrated   that appropriate dark matter-baryon interactions can
accommodate neutron stars with mass above two solar masses. In the same direction lies the finding of Ref.~\cite{McKeen-2018} according to which the existence of neutron stars with a mass greater than  $0.7 M_{\odot}$ places severe
constraints on such particles, requiring them to be heavier than 1.2 GeV or to have strongly repulsive
self-interactions.

In the present work we consider the decay channel  
$
    n\longrightarrow  \chi +\phi
$
where $\chi$ is the dark matter particle and  $\phi$ is a dark boson that has a very small mass.  It is worth mentioning here that other additional mechanisms of decay have been proposed. These mechanisms are $n\longrightarrow  \chi +e^+ +e^-$ and $n\longrightarrow  \chi +\gamma$. However, the above mechanisms have been ruled out by  experimental measurements~\cite{Tang-2018,Sun-2018}.

It is important to emphasize that the mechanisms described above have thus far only been studied under laboratory conditions. In contrast, the extreme environment within neutron stars, characterized by vastly different pressures, densities, temperatures, and surrounding matter could significantly alter the feasibility of neutron dark decay. These conditions might either enhance or suppress the process entirely. As such, this represents an additional and critical parameter in the problem, one that cannot be ignored, yet currently cannot be resolved with scientific certainty.

Until very recently, NSs or compact objects with mass below the canonical mass of $1.4M_{\odot}$ were unclassified. Specifically, within the last year the compact object HESS J1731-347 was reported with a mass of $M=0.77^{+0.20}_{-0.17}~{\rm M_{\odot}}$ and a radius of $R=10.4^{+0.86}_{-0.78}$ km~\cite{HESS-2023}, while a new analyses of the compact object XTE J1814-338~\cite{Kini-2024, Baglio-2013} revealed a mass of $M=1.2^{+0.05}_{-0.05}~{\rm M_{\odot}}$ and a radius of $R=7^{+0.4}_{-0.4}$ km. In view of the above, several light compact objects were reported which can hardly be explained simultaneously and at the same time reproduce the NS maximum mass constraint. Furthermore, besides the HESS J1731-347 and XTE J1814-338 objects, also a new analysis of PSR J1231-1411~\cite{Salmi-2024} ($M=1.04^{+0.05}_{-0.03}~{\rm M_{\odot}}$ and radius $R=12.6^{+0.3}_{-0.3}$ km) appeared recently, noting that the authors report problems related to the convergence in the fitting procedure and the uncertainty might be larger than reported. Such variety of light compact objects obviously requires variety also in physics scenarios, characterized by different structure. 
 
Considerable research has been dedicated to interpreting HESS
J1731-347 as either normal NS or a hybrid star. In  particular, in Ref.~\cite{Veselsky-2025a} the authors found that the onset of kaon condensate accurately describes the ultralight compact object within the supernova remnant HESS J1731-347. Moreover, 
its properties have been studied within the RMF model by Kubis et al.~\cite{Kubis-2023} and
within the mean field model with a parity doublet structure by Gao et
al.~\cite{Gao-2024}. Hybrid stars comprising quarks or heavy baryons alongside nucleons have also been explored in Refs.~\cite{Li-2023,Brodie-2023,Mariani-2024,Li-2024}. Additionally, Sagun et
al.~\cite{Sagun-2023} investigated various scenarios in which HESS J1731-347 could
be described as a NS, a hybrid star, a strange star, or a star with an
admixture of dark matter

Several studies attempt to reproduce the properties of the XTE J1814-338 using a concept of hybrid star. Specifically, Pitz and Schaffner-Bielich \cite{Pitz-2024} studied the possibility of the XTE J1814-338 being a bosonic star with a nuclear matter core, while Yang \textit{et al.}~\cite{Yang-2024} investigated the scenario of a strange star admixed with mirror dark matter. Recently, Lopes and Issifu~\cite{Lopes-2024} claimed that  this object can be explained also as a dark matter admixed neutron star. Notably, the idea of ultra compact stars has been studied by Li \textit{et al.}~\cite{Sedrakian-2023} before the results of Ref.~\cite{HESS-2023}. 
Recent work~\cite{Laskos-2024} achieves the agreement with the XTE J1814-338 by applying an assumption of first order phase transition with specifically chosen transition density and energy density jump. 
Another possibility is the existence of strangeness, which can be considered
as an additional degree of freedom, the amount of which might be determined by the formation scenario of a specific object. In Ref.~\cite{Veselsky-2025b} the authors managed  to explain the bulk properties of the XTE J1814-338 object and at the same time the HESS J1731-347 object, using a mixture of kaon condensation in dense nuclear matter. Moreover, they proposed  that to  simultaneously explain the current variety of astrophysical objects, it is essential to resurrect a scenario of two distinct branches, each corresponding to a different composition of nuclear matter.
Thus, compact objects with strangeness can be produced in collapse of more massive stellar objects, where higher temperatures and densities allow copious production of strangeness~\cite{Haensel_2007, Brown-Rho_2008}.

The primary motivation of the present study is to investigate to what extent the recently observed exotic compact objects are compatible with the recently proposed hypothesis of neutron dark decay. We have strong reasons to believe that these objects cannot be accounted for, at least with the required level of reliability, by purely hadronic equations of state. Additional assumptions are therefore necessary to construct equations of state that are compatible with such objects. The central parameters of this investigation are the strengths of the interactions among the dark particles and between the dark particles and neutrons. We find that these parameters play a decisive role in predicting the existence and properties of these exotic compact objects. The results and conclusions of the present study are valuable not only because they provide an estimate of the impact of dark matter on neutron star properties, but also because they enable observationally informed constraints to be placed on the strengths of the relevant interactions. In any case, the extent to which the proposed dark decay of the neutron is influenced by the extreme conditions prevailing in neutron stars remains uncertain. Within the framework of the preceding consideration,  we additionally examined the scenario in which a mechanism suppressing the dark neutron decay becomes operative at densities  above nuclear saturation density. Remarkably, we find that this mechanism yields stellar configurations that reproduce the observational data of HESS J1731-347 with high fidelity, while simultaneously predicting maximum masses that comfortably exceed the two solar mass thresholds.

The paper is organized as follows: In Section II, we briefly present the concept of  neutron dark  decay in neutron stars. Section III is dedicated to the construction of the equation of state, while Section IV presents and discusses the results. Finally, Section V concludes the study with our final remarks.

\section{Neutron dark  decay}
The dominant neutron decay channel is the classical $\beta$-decay where
\begin{equation}
n\longrightarrow p+e^{-}+\bar{\nu}_e
\label{beta-1}
\end{equation}
In the present work, following the suggestion in Ref.~\cite{Fornal-2018}, we employ an additional mechanism, concerning the neutron decay according to (see also Refs.~\cite{Gil-2024,Motta-2018a,Motta-2018b,Shirke-2023,Shirke-2024})
\begin{equation}
n\longrightarrow \chi+\phi,
\label{beta-2}
\end{equation}
where $\chi$ is a dark spin-1/2 fermion with baryon number 1, and $\phi$ is a light dark boson.   Following the analysis of Ref.~\cite{Fornal-2023} there are some restrictions on the range of dark particle masses. In order to allow for the neutron decay channel into $\phi$ while ensuring the stability of nuclei, one must impose~\cite{Fornal-2023,Gil-2024} 
\begin{equation}
  937.993 \ {\rm MeV} < m_{\chi}+m_{\phi} < 939.565 \ {\rm MeV}.
  \label{mass-range}
\end{equation}
Moreover, the mass of the DM particle is bounded from below, $m_{\chi}>937.993$ MeV to prevent the decay of $^9$Be
triggered by the neutron dark decay. 
The final state fermion $\chi$ and  the scalar $\phi$ can be dark matter candidates if they  are stable, and this condition is ensured  when~\cite{Fornal-2023} 
\begin{equation}
|m_{\chi}-m_{\phi}|<m_p+m_e=938.783 \ {\rm MeV}.
\label{cond-2}
\end{equation}
We consider that in neutron stars, the presence of the light dark boson $\phi$ is completely irrelevant since it  escapes from the neutron star interior very quickly and  therefore does not participate in the construction of the equation of state of  the star. Finally,  we fix the  particle mass at $m_{\chi}=938$ MeV~\cite{Motta-2018a,Motta-2018b}.

\section{Equation of state}
\subsection{Hadronic Equation of State}
In  the present work, in order to describe the neutron star matter we use the Equation of State (EoS) derived by  Akmal et al.~\cite{Akmal-1998} (hereafter APR) in particular the model  A18+UIX (for pure neutron matter). This is an EoS with microscopic origin and its predictions are in very good agreement with both the measured maximum masses (see PSR J1614-2230~\cite{Arzoumanian-2018}, PSR J0348+0432~\cite{Antoniadis-2013}, PSR J0740+6620~\cite{Cromartie-2020}, 
and PSR J0952-0607~\cite{Romani-2022} pulsar observations for the possible maximum mass) and some astrophysical constraints for radii (see the GW170817 event~\cite{Abbott-2019-X}). It is worth to notice here that the present study mainly focuses on the EoS of DM and its properties. Nevertheless, the use of a realistic EoS for neutron stars provides a guarantee for the reliability of our results.


\subsection{Dark matter equation of state }
Regarding the DM particles, we assume they are  fermions that interact with each other through a repulsive force.
 We consider a Yukawa-type interaction for this purpose (see also Refs.~\cite{Gil-2024,Shirke-2023,Shirke-2024,Cline-2018,Nelson-2019})
\begin{equation}
V(r)=\frac{{\rm g}_{\chi}^2 (\hbar c)}{4\pi r} \exp\left[-\frac{m_Vc^2}{\hbar c}r\right]
\label{Yukawa-1}
\end{equation}
where ${\rm g}_{\chi}$  and $m_{V}$ are   the coupling constant  and  the mediator mass, respectively. The contribution to the energy density of the self-interaction is given by
\begin{equation}
{\cal E}_{SI}(n_{\chi})=\frac{(\hbar c)^3}{2z_{\chi}^2} n_{\chi}^2  
\end{equation}
where $z_{\chi}=m_{V}c^2/{\rm g}_{\chi}$ (in units MeV).
The total energy density of the dark matter particles ${\cal E}_{\chi}(n_{\chi})$ is given by
\begin{eqnarray}
{\cal E}_{\chi}(n_{\chi})&=&\frac{(m_{\chi}c^2)^4}{(\hbar c)^38\pi^2}\left[x\sqrt{1+x^2}(1+2x^2)\right.\nonumber \\
&-&\left.\ln(x+\sqrt{1+x^2})    \right]
+ \frac{n_{\chi}^2(\hbar c)^3}{2z_{\chi}^2}
\label{Kov-ex}
\end{eqnarray} 
where
\[x=\frac{(\hbar c)(3\pi^2n_{\chi})^{1/3}}{m_{\chi}c^2} \]
The  parameter $z_{\chi}$ is treated as a free parameter related to the strength of the self-interaction. For comparison to similar studies this parameter is related to the interaction strength $G_{\chi}$  by
$z_{\chi}=\hbar c/\sqrt{G_{\chi}}$~\cite{Das-Burgio-2025}.
Also, it is worth mentioning here that there exist astronomical constraints on the values of $G_{\chi}$, with typical values 4-135 fm$^{2}$ \cite{Das-65,Das-66,Das-67,Das-68,Das-69}, which correspond to the values used in the present study (for a detailed analysis see Ref.~\cite{Das-Burgio-2025}). The only exception is the value $z_{\chi}=10$ MeV which refers to a very strong interaction and can be considered  an upper limit in the present study. 

As we mentioned before, the DM is a self-interacting Fermi gas, where the contribution on the energy density of the self-interaction is given by $\frac{y^2}{2} (\hbar c)^3 n_{\chi}^2$, where $n_{\chi}$ is the DM density and $y={\rm g}_{\chi}/m_{\phi}c^2$ (in units MeV$^{-1}$), is the interaction strength. The interaction strength $y$ can be constrained by observational limits. These limits imposed on the cross section of the self-interaction~\cite{Burgio-2024,Liu-2024b}. In particular, according to \cite{Mark-2004,Das-66,Das-67,Loeb-2022} it holds $\sigma/ m_{\chi}\sim 0.1-10 \ {\rm cm}^2 /{\rm g}$. Moreover,  it has been showed that  the Born approximation is very accurate in the region $m_{\chi} \sim 1 \ {\rm GeV}~$\cite{Kouvaris-15,Maselli-2017,Tulin-2013}. Thus, we found that the interaction parameter $y$  varies in the range $\sim (0.001-0.1)({\rm GeV}/ m_{\chi}c^2)^{1/4} ({\rm MeV}^{-1})$~\cite{Burgio-2024}.

In order to enrich our study, we consider also the case of a repulsive  effective
potential between the neutrons and the DM particles  through the
exchange of the light scalar $\tilde{\phi}$ and  a Yukawa-type interaction (for more details see also Refs.~\cite{Gil-2024,Grinstein-2019})
\begin{equation}
V(r)=\frac{{\rm \tilde{g}}_{\chi} {\rm g}_{n}(\hbar c)}{4\pi r} \exp\left[-\frac{m_{\tilde\phi}c^2}{\hbar c}r\right]
\label{Yukawa-1}
\end{equation}
where ${\rm \tilde{g}}_{\chi}$  and $m_{\tilde{\phi}}$ are   the coupling constant  and  the mediator mass respectively.  The contribution to the energy density due to the neutron-DM interaction is given by
\begin{equation}
{\cal E}_{\rm int}(n_n, n_{\chi})=\frac{n_{\chi}n_n(\hbar c)^3}{2z_{\chi n}^2}, 
\end{equation}
where $z_{\chi n}\equiv m_{\tilde{\phi}}c^2/\sqrt{|{\rm \tilde{g}}_{\chi}{\rm g}_{n}|}$.  The coupling constants $g_{\chi}, g_n$ are dimensionless.  
The parameter ranges for $z_{\chi n}$ are constrained and thoroughly discussed in Ref.~\cite{Gil-2024}, taking into account both the bulk properties of neutron stars (such as mass and radius) and experimental data related to neutron beta decay.

\subsection{The total equation of state }
By combining the equations of state of neutron matter and dark matter, we obtain the total energy density  ${\cal E}_{\rm tot}(n_n,n_{\chi})$, which can be written as 
\begin{equation}
{\cal E}_{\rm tot}(n_n,n_{\chi})={\cal E}_{\rm nucl}(n_n)+{\cal E}_{\chi}(n_{\chi})+{\cal E}_{\rm int}(n_n,n_{\chi})
\label{Ener-tot}
\end{equation}
The minimization of the total energy density ${\cal E}_{\rm tot}(n_n,n_{\chi})$ with respect to the number density $n_{\chi}$ of DM  or neutrons $n_n$ (for a fixed total density $n_{\rm tot}\equiv n=n_{\chi}+n_n$) lead to the equilibrium condition $\mu_{\chi}=\mu_n$. Now, since by definition, we have 
\[ \mu_{\chi}=\frac{\partial {\cal E}_{\rm tot}(n_n,n_{\chi})}{\partial n_{\chi}},\quad   \mu_{n}=\frac{\partial {\cal E}_{\rm tot}(n_n,n_{\chi})}{\partial n_{n}}   \]
the mentioned equilibrium condition  leads  to the equation
\begin{equation}
\frac{\partial {\cal E}_{\rm nucl}(n_n)}{\partial n_n}-\frac{\partial  {\cal E}_{\chi}(n_{\chi})}{\partial n_{\chi}}+
\frac{(\hbar c)^3}{2z_{\chi n}^2}\left(n_{\chi}-n_n\right)=0
\label{chem-1}
\end{equation}
Eq.~(\ref{chem-1})  
leads to the calculation of the threshold of creation of the dark matter particle with mass $m_{\chi}$ as well as on the relation between the $n_{\chi}$  and $n_n$. In particular,  Eq.~(\ref{chem-1}) must be solved numerically  (for fixed value of $m_{\chi}$, $z_{\chi}$ and/or $z_{\chi n}$  in order to find the dependence $n_{\chi}=n_{\chi}(n_n)$ or even better $\chi=\chi(n)$, where $n=n_{\chi}+n_n$ and  $\chi=n_{\chi}/n$.
In the case without interaction between the neutron and DM Eq.~(\ref{chem-1}) is reduced to the form
\begin{equation}
\frac{\partial  {\cal E}_{\rm nucl}(n_n)}{\partial n_n}=\frac{\partial  {\cal E}_{\chi}(n_{\chi})}{\partial n_{\chi}}
\label{chem-2}
\end{equation}
Finally, the total pressure $P_{\rm tot}(n_n,n_{\chi})$ takes the  form
\begin{equation}
P_{\rm tot}(n_n,n_{\chi})=n_n\mu_n+n_{\chi}\mu_{\chi}- {\cal E}_{\rm tot}(n_{\chi},n_n)
\label{Pres-tot}
\end{equation}%
Eqs.~(\ref{Ener-tot}) and  (\ref{Pres-tot})  jointly define the equation of state of mixed neutron and dark matter and constitute the primary input for the calculation of the fundamental properties of the corresponding compact objects.

\subsection{TOV equations and dark  matter fraction }
The bulk properties of neutron stars, and, by extension, of compact objects formed from an admixture of neutron matter and dark matter, such as their mass and radius, are determined by solving the coupled Tolman–Oppenheimer–Volkoff (TOV) equations.~\cite{Shapiro:1983du,Haensel2007NeutronS1,schaffner-bielich_2020}. 
This system reads
\begin{eqnarray}
\frac{dP(r)}{dr}&=&-\frac{G{\cal E}(r) M(r)}{c^2r^2}\left(1+\frac{P(r)}{{\cal E}(r)}\right) \nonumber \\
&\times&
 \left(1+\frac{4\pi P(r) r^3}{M(r)c^2}\right) \left(1-\frac{2GM(r)}{c^2r}\right)^{-1},
\label{TOV-1}
\end{eqnarray}
\begin{equation}
\frac{dM(r)}{dr}=\frac{4\pi r^2}{c^2}{\cal E}(r).
\label{TOV-2}
\end{equation}
As discussed in the preceding section, the equation of state ${\cal E}={\cal E}(P)$, serving as the primary input for the TOV equations, is constructed by simultaneously incorporating the contributions of neutrons and dark matter particles.

A key parameter characterizing the impact of dark matter on the stellar mass is the dark matter fraction, defined as the ratio of the dark-matter contribution to the total mass of the star. Specifically, this fraction is given by (see also Refs.~\cite{Shirke-2023,Shirke-2024})
\begin{equation}
f_{\chi}\equiv \frac{M_{\chi}}{M}=
\frac{\int {\cal E}_{\chi} d^3r}{\int {\cal E}_{\rm tot} d^3r}
\label{fraction}
\end{equation}
It is important to clarify that, whereas the energy density of dark matter particles is well defined when only self-interactions are considered, this is no longer the case when interactions with neutrons are included, owing to the additional term arising from their coupling to the neutron component. In this case, we assume that the contribution to the total mass is determined based on the relative fraction of each particle species. Thus, the term
\begin{equation}
{\cal E}_{\rm int}^{\rm DM}=  \frac{n_{\chi}}{n_n+n_{\chi}} \frac{n_{\chi}n_n(\hbar c)^3}{z_{\chi n}^2}
\label{dm-mass}
\end{equation}
accounts for the contribution to the dark-matter component, while the term
\begin{equation}
{\cal E}_{\rm int}^{ n}=\frac{n_{n}}{n_n+n_{\chi}} \frac{n_{\chi}n_n(\hbar c)^3}{z_{\chi n}^2}
\label{DM=mass}
\end{equation}
corresponds to the contributions of neutrons.

\begin{figure*}[ht]
\centering
\includegraphics[width=0.32\textwidth]{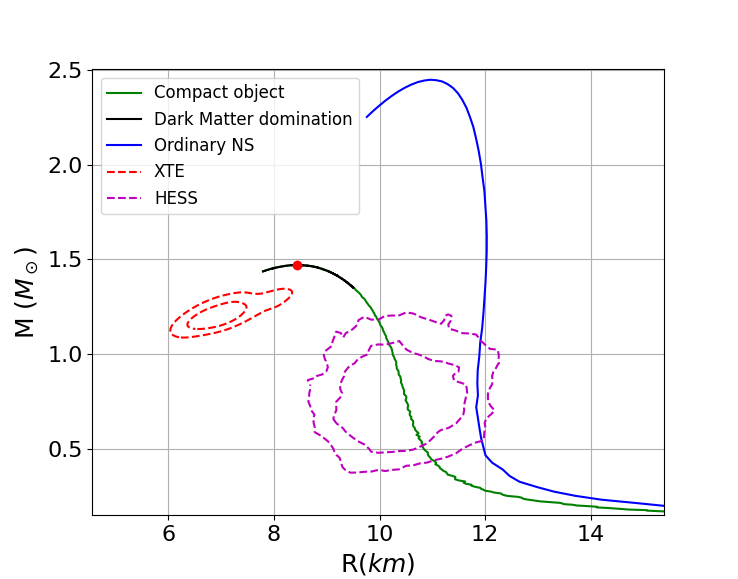}
\includegraphics[width=0.32\textwidth]{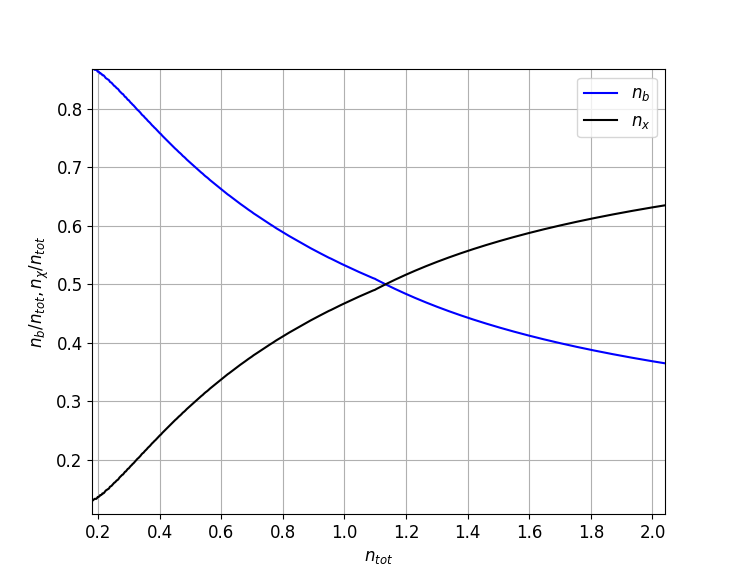}
\includegraphics[width=0.32\textwidth]{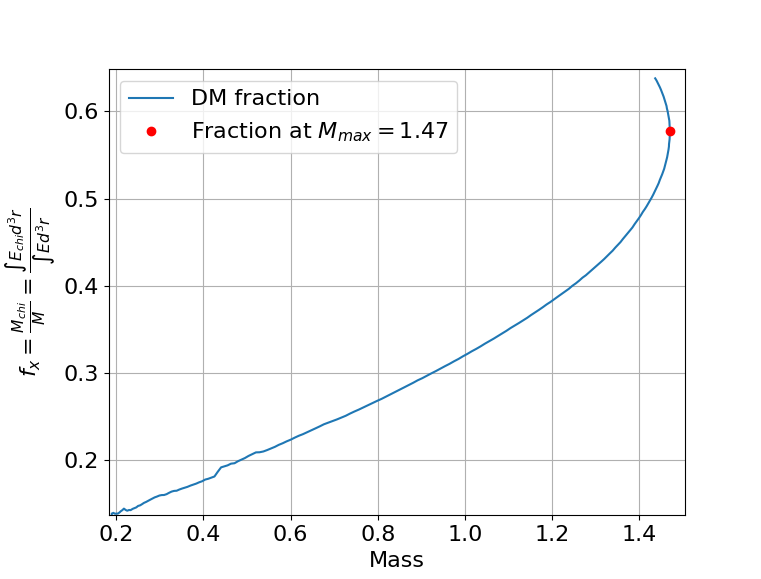}
\includegraphics[width=0.32\textwidth]{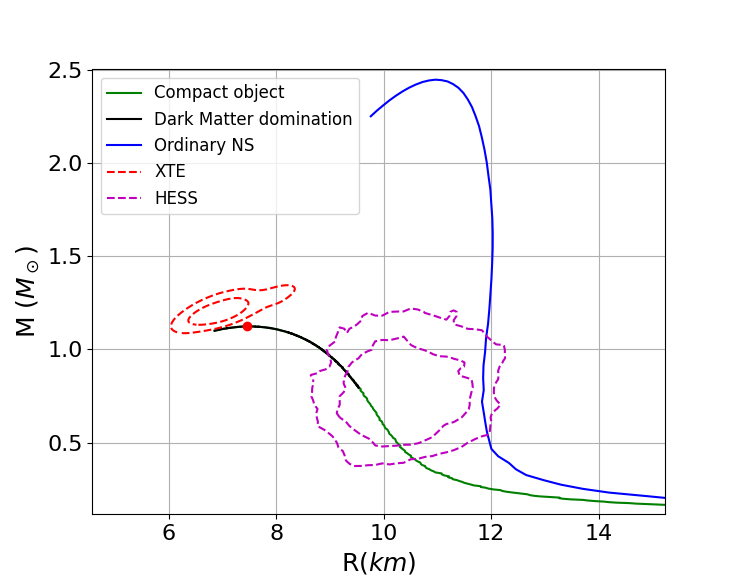}
\includegraphics[width=0.32\textwidth]{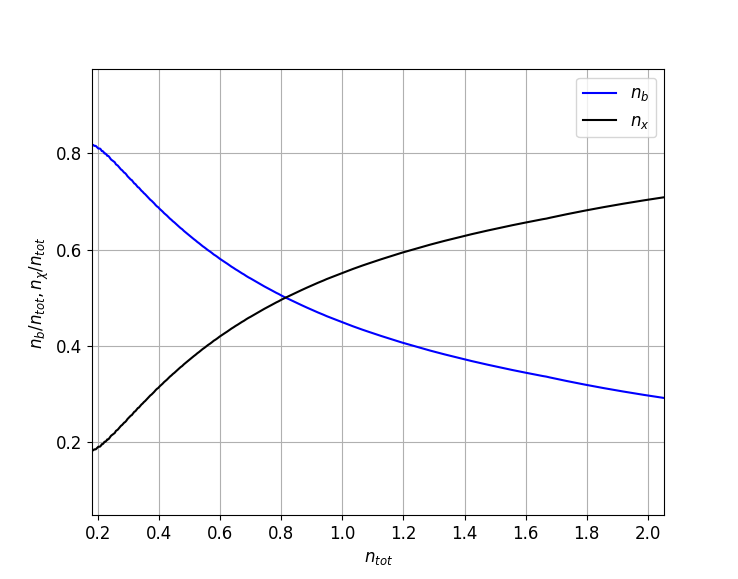}
\includegraphics[width=0.32\textwidth]{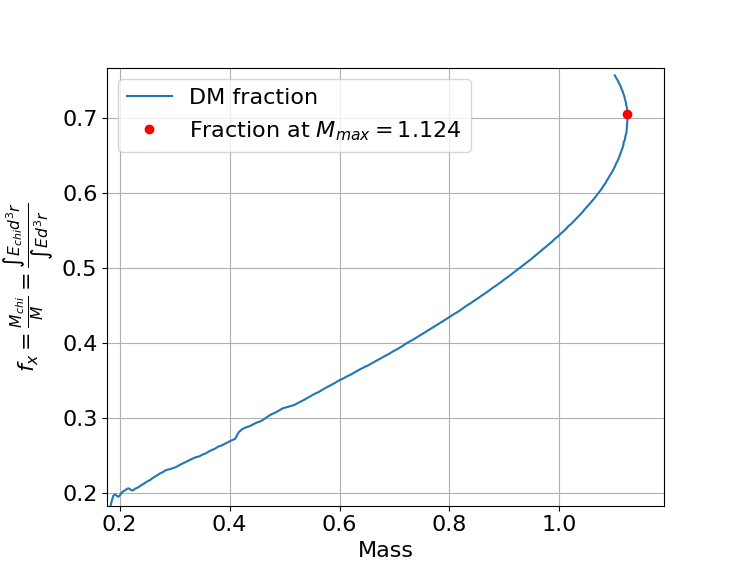}
\includegraphics[width=0.32\textwidth]{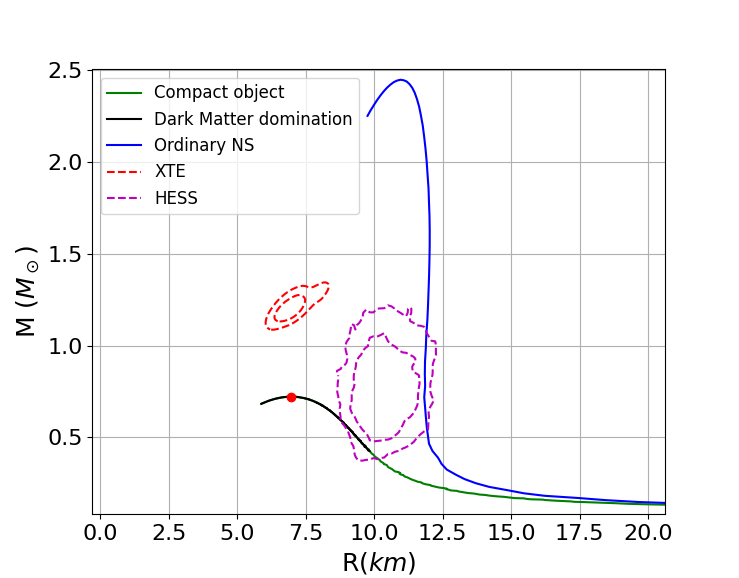}
\includegraphics[width=0.32\textwidth]{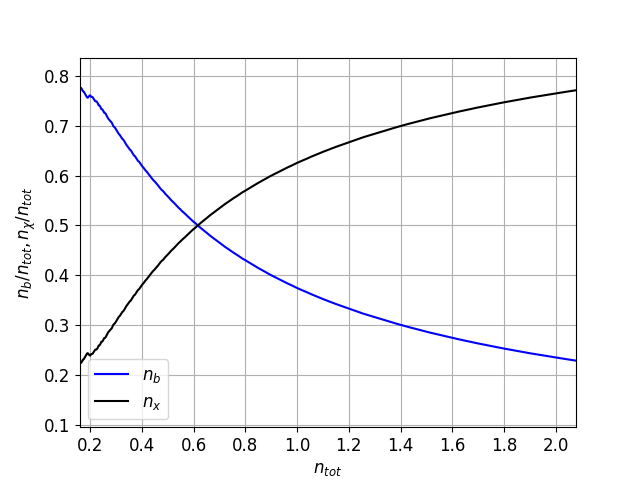}
\includegraphics[width=0.32\textwidth]{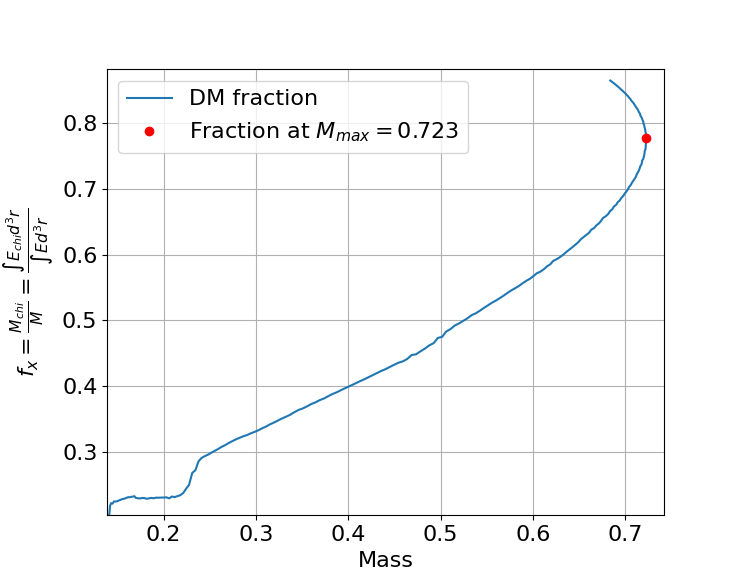}
\caption{(a)  The ${\rm M-R}$ diagram (for the cases of  the ordinary neutron star matter (Ordinary NS) and for the dark matter-neutron matter admixture (Compact object)). Dark matter domination refers to the mass range in which the dark particle fractions  exceeds that of neutron fractions in the center of the compact object. (b) The number density fractions and (c) the mass fraction $f_{
\chi}$ for the self-interaction cases  $z_{\chi}=130,220,1500$ MeV (from the top to the bottom). The observational constraints for the HESS J1731-347~\cite{HESS-2023} and for the  XTE J1814-338~\cite{Kini-2024, Baglio-2013}  are also indicated.  }
 \label{zx-only}
\end{figure*}


\begin{figure*}[ht]
\centering
\includegraphics[width=0.32\textwidth]{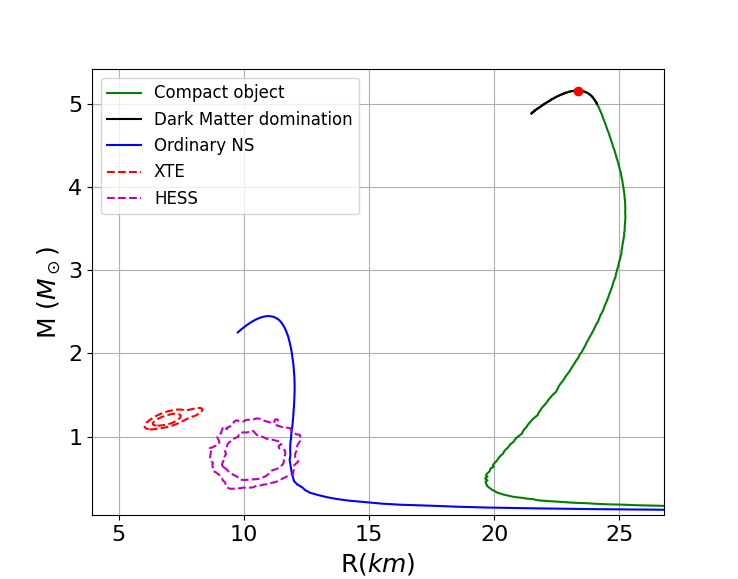}
\includegraphics[width=0.32\textwidth]{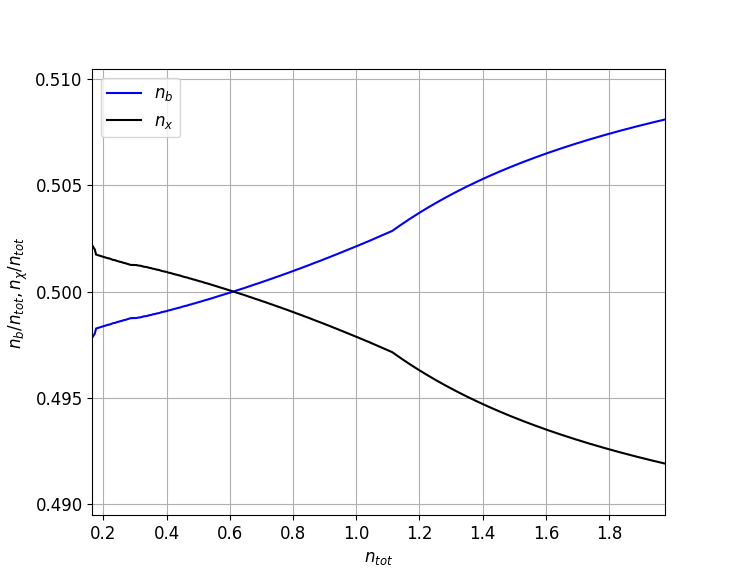}
\includegraphics[width=0.32\textwidth]{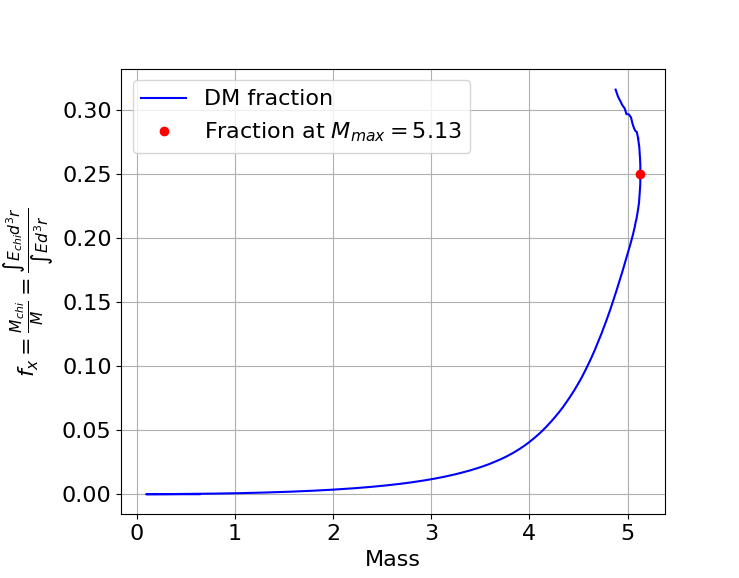}
\includegraphics[width=0.32\textwidth]{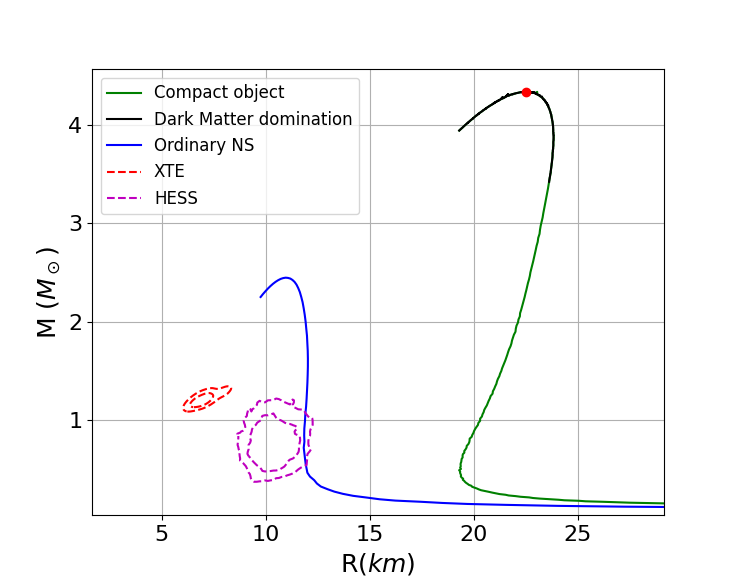}
\includegraphics[width=0.32\textwidth]{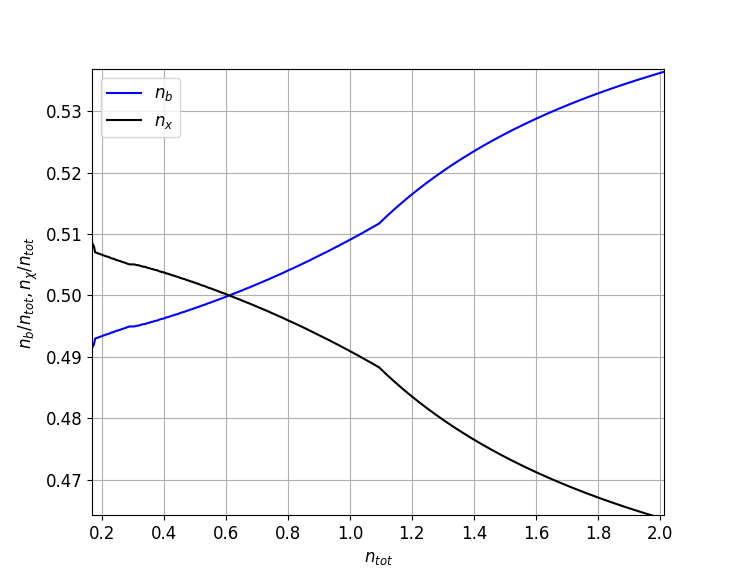}
\includegraphics[width=0.32\textwidth]{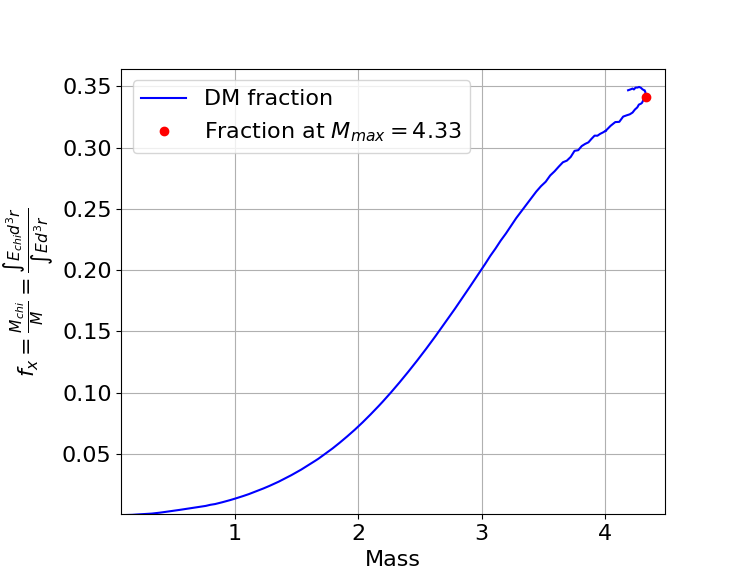}
\includegraphics[width=0.32\textwidth]{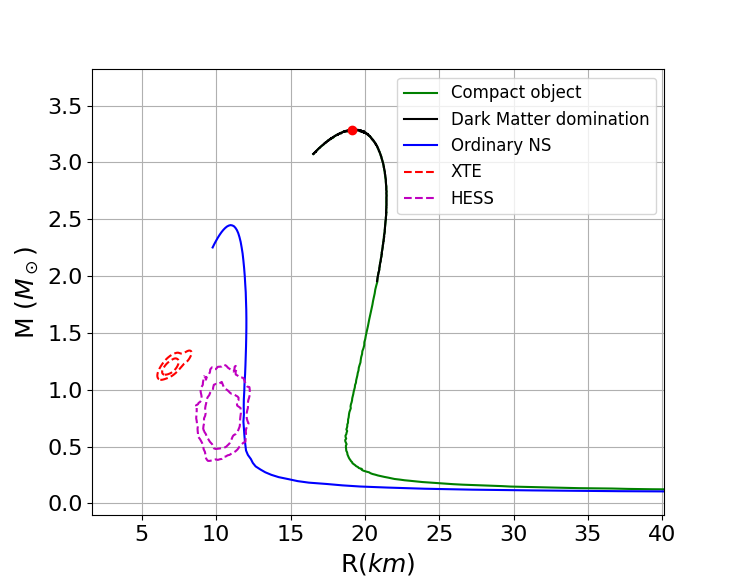}
\includegraphics[width=0.32\textwidth]{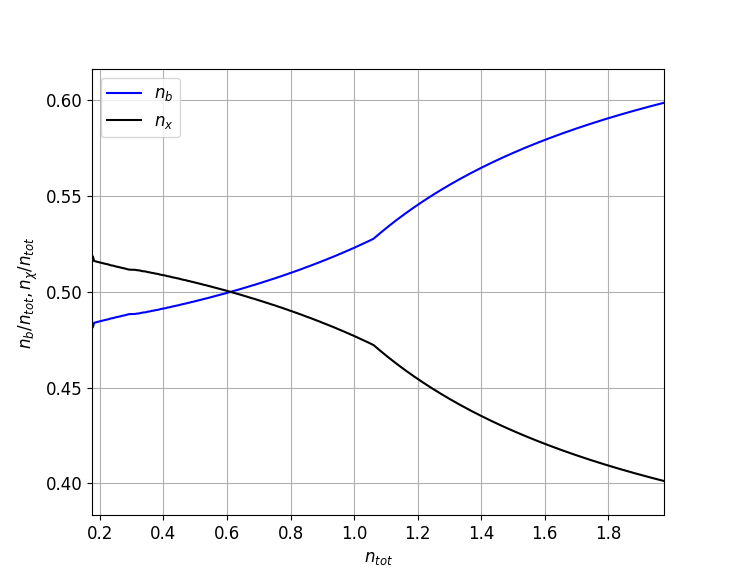}
\includegraphics[width=0.32\textwidth]{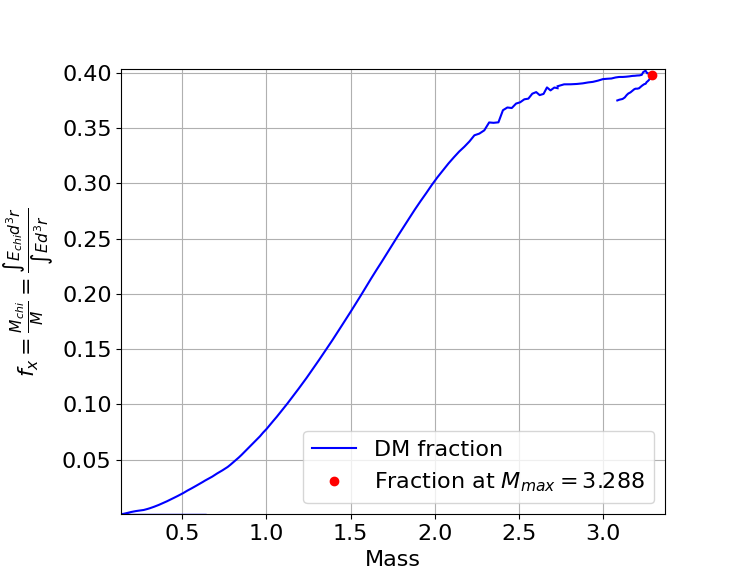}
\includegraphics[width=0.32\textwidth]{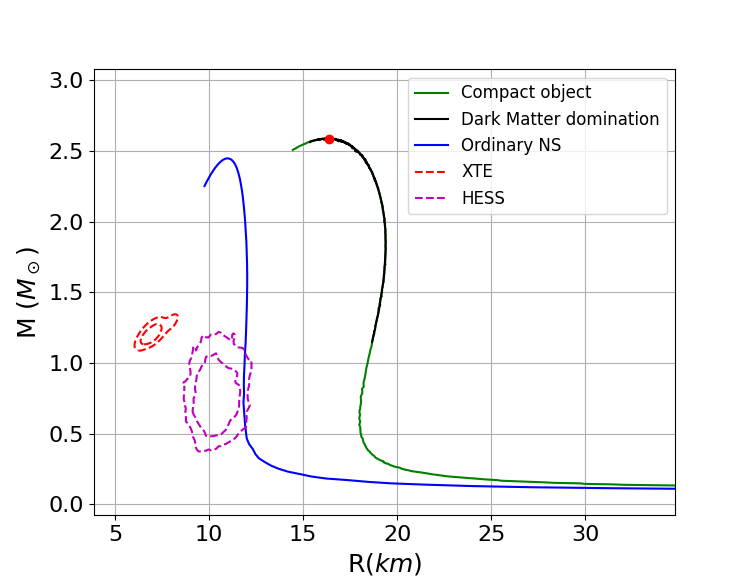}
\includegraphics[width=0.32\textwidth]{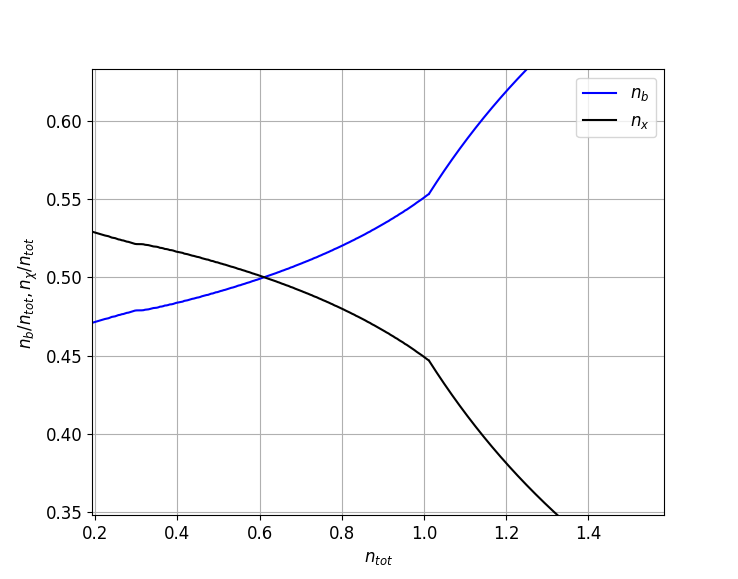}
\includegraphics[width=0.32\textwidth]{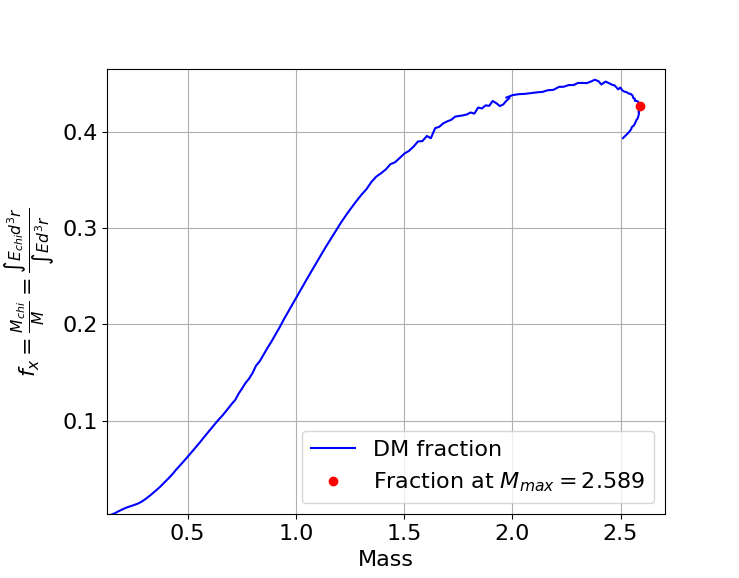}
\caption{The same as the Fig.~(\ref{zx-only}) for  $z_{\chi n}=10, 20, 30,40$ MeV (from the top to the bottom). }
   \label{znx-only}
\end{figure*}


\begin{figure*}[ht]
\centering
\includegraphics[width=0.32\textwidth]{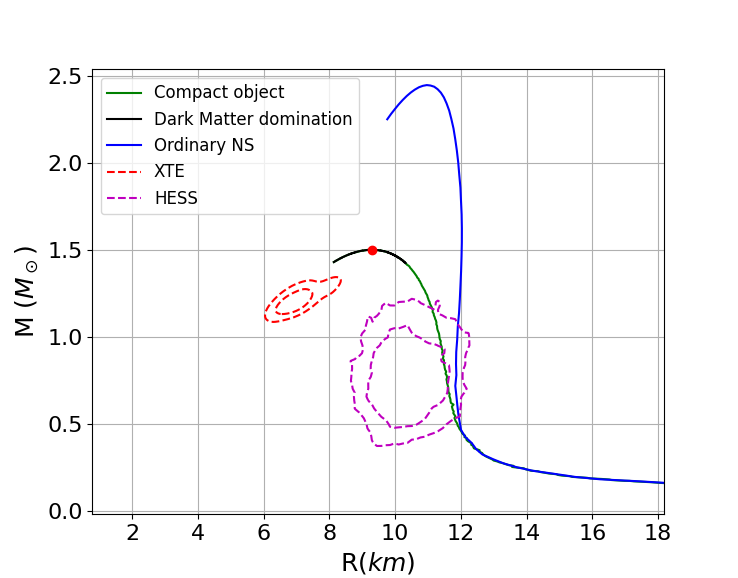}
\includegraphics[width=0.32\textwidth]{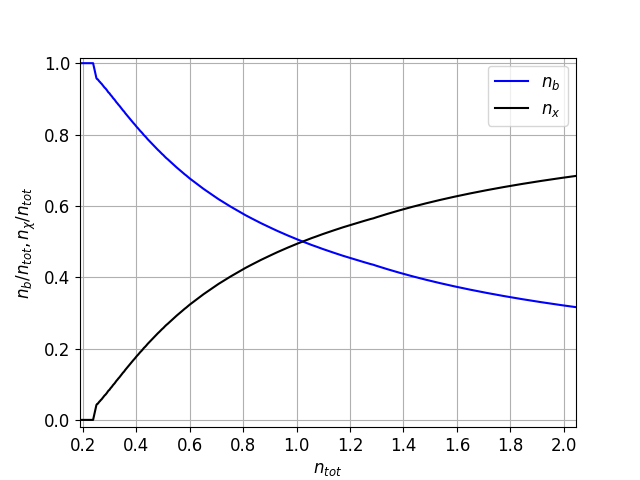}
\includegraphics[width=0.32\textwidth]{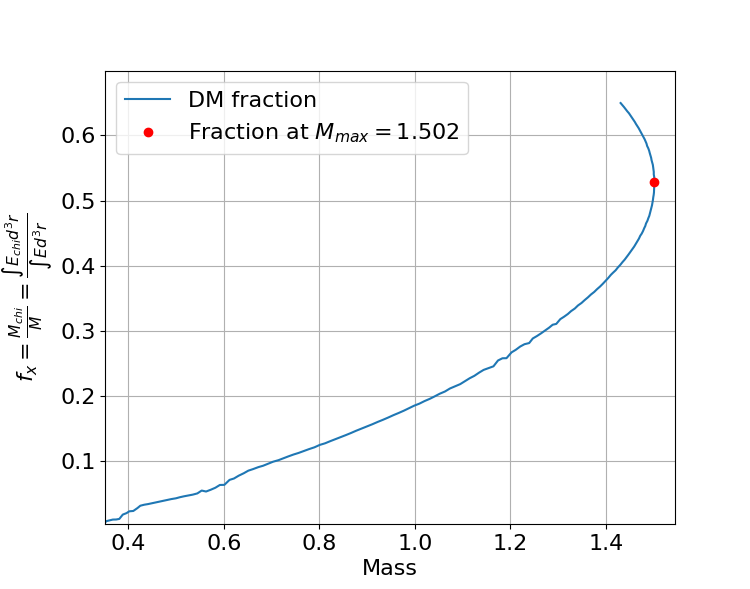}
\includegraphics[width=0.32\textwidth]{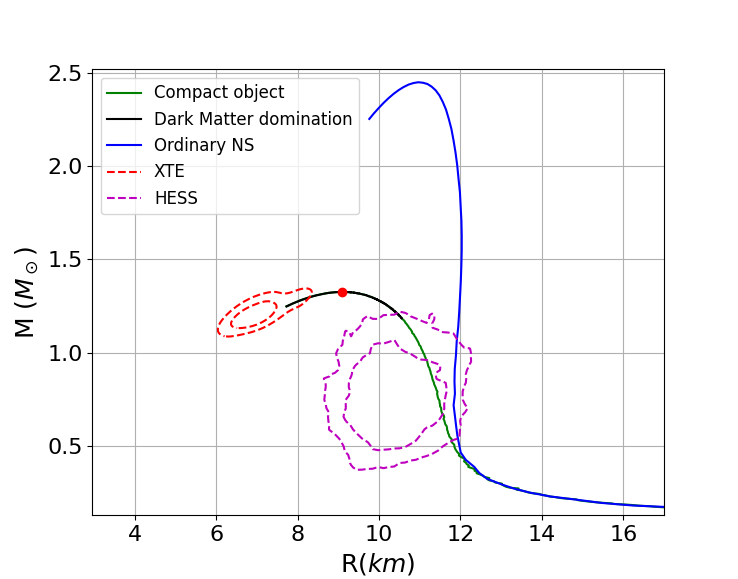}
\includegraphics[width=0.32\textwidth]{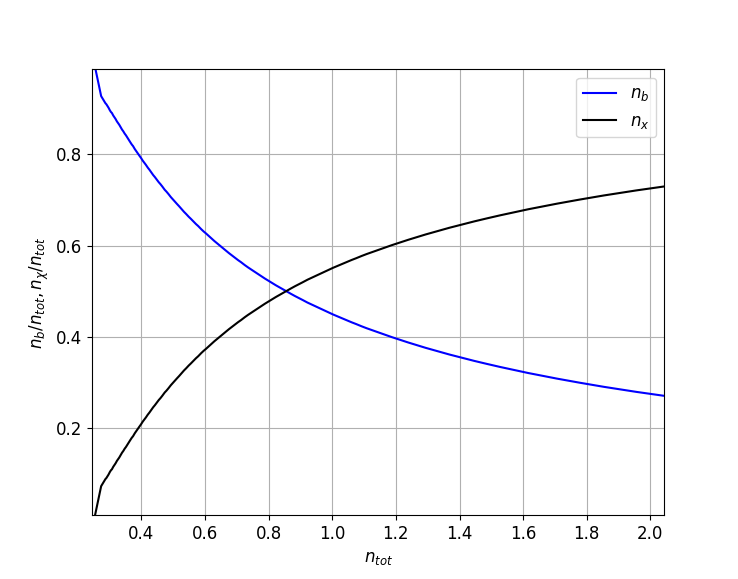}
\includegraphics[width=0.32\textwidth]{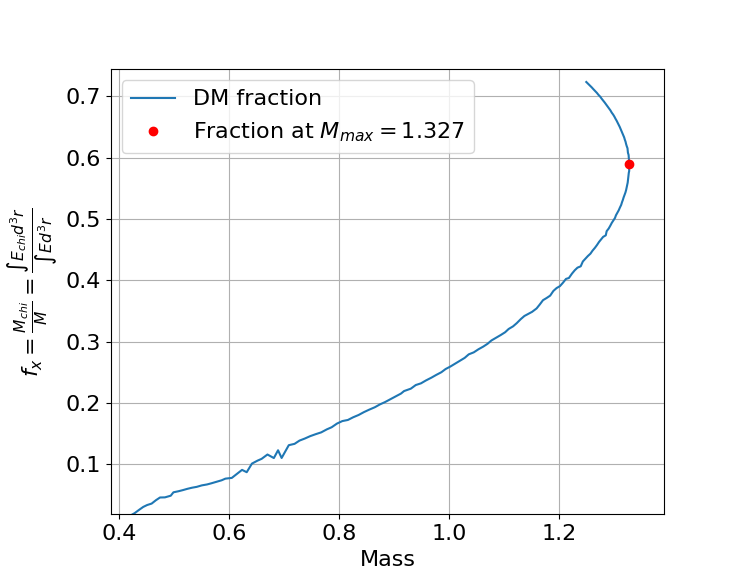}
\includegraphics[width=0.32\textwidth]{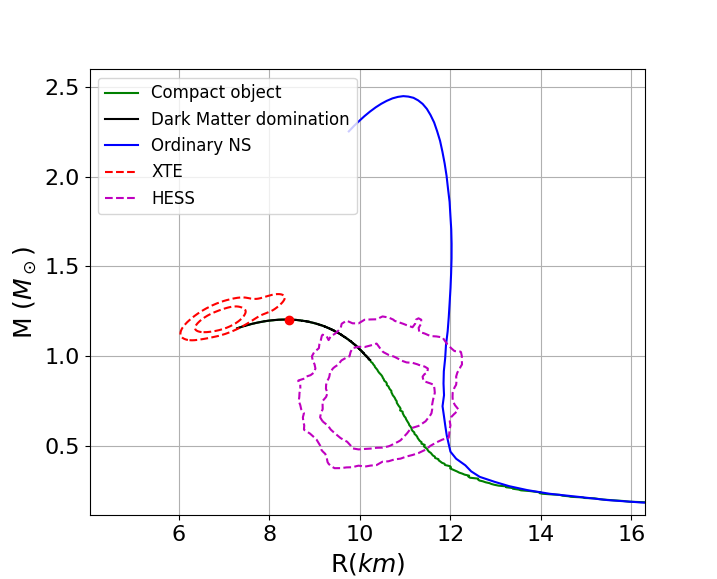}
\includegraphics[width=0.32\textwidth]{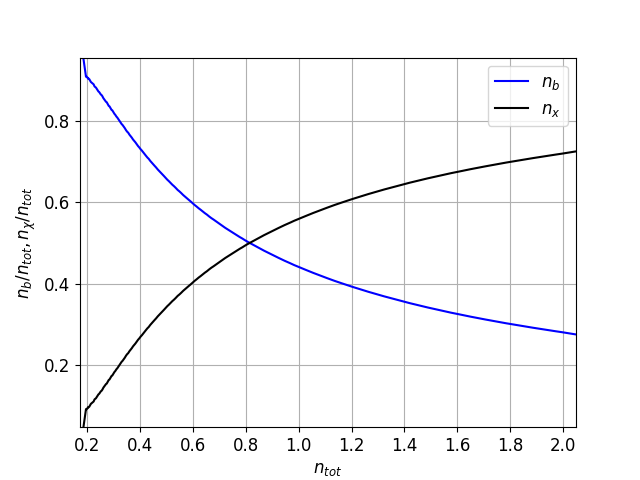}
\includegraphics[width=0.32\textwidth]{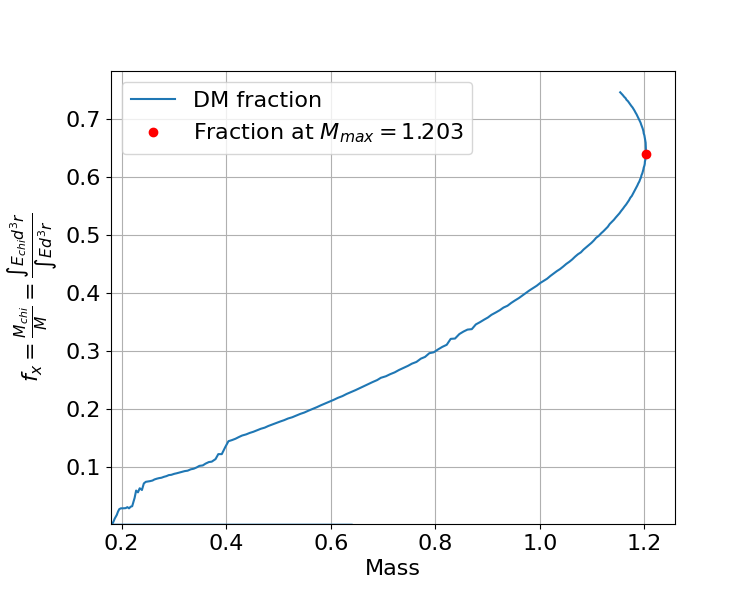}
\includegraphics[width=0.32\textwidth]{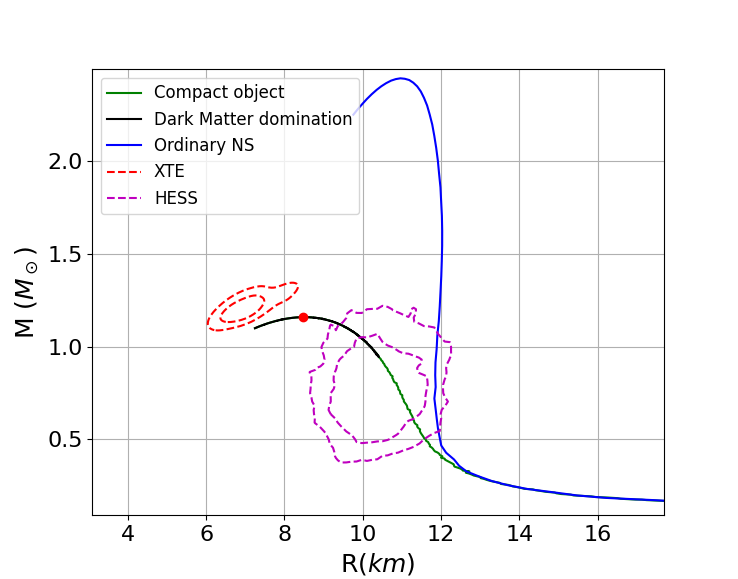}
\includegraphics[width=0.32\textwidth]{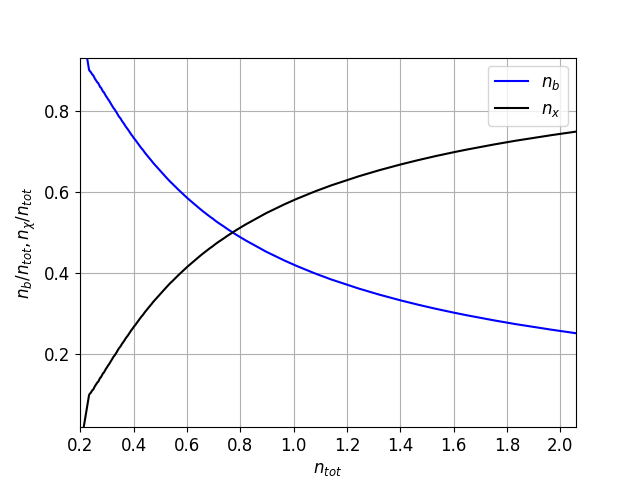}
\includegraphics[width=0.32\textwidth]{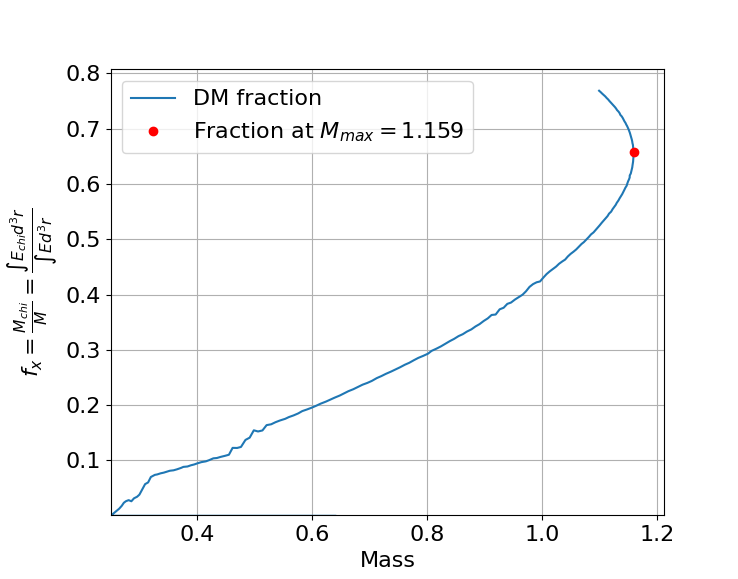}
\includegraphics[width=0.32\textwidth]{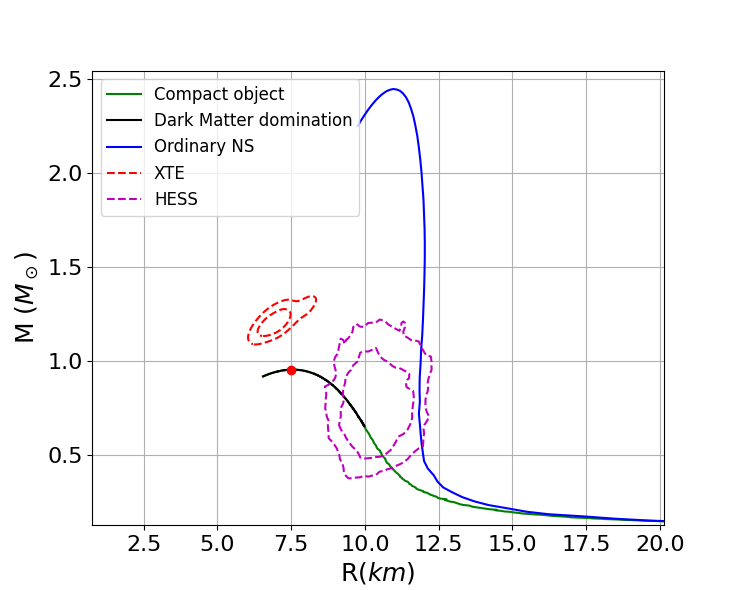}
\includegraphics[width=0.32\textwidth]{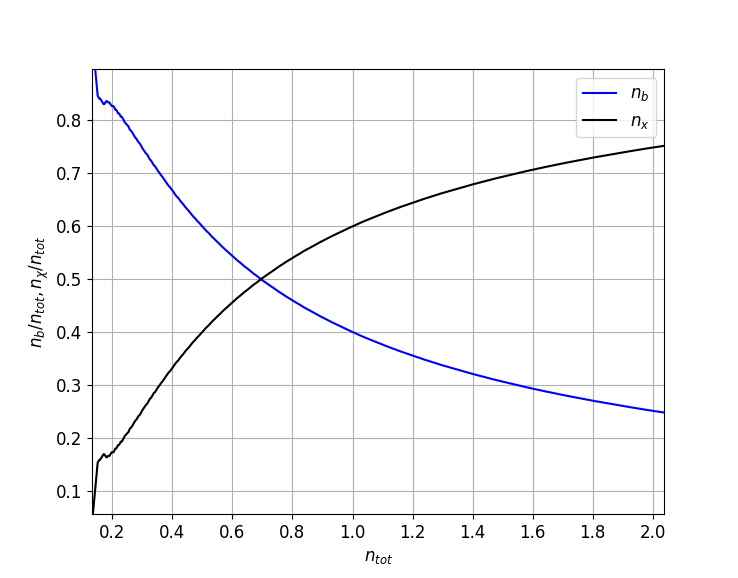}
\includegraphics[width=0.32\textwidth]{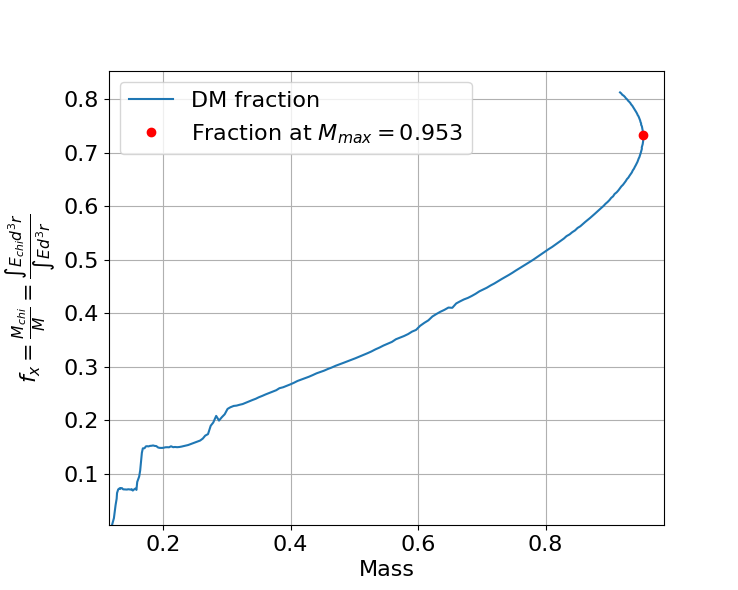}
\caption{The same as the Fig.~(\ref{zx-only}) for a)  $z_{\chi}=150$ MeV and  $z_{\chi n}=150$ MeV, b) $z_{\chi}=200$ MeV and  $z_{\chi n}=150$ MeV, c) $z_{\chi}=220$ MeV and  $z_{\chi n}=200$ MeV, 
d) $z_{\chi}=250$ MeV and  $z_{\chi n}=180$ MeV,
e) $z_{\chi}=350$ MeV and  $z_{\chi n}=300$ MeV (from the top to the bottom).  }
\label{zx-zxn}
\end{figure*}


\begin{figure*}[ht]
\centering
\includegraphics[width=0.44\textwidth]{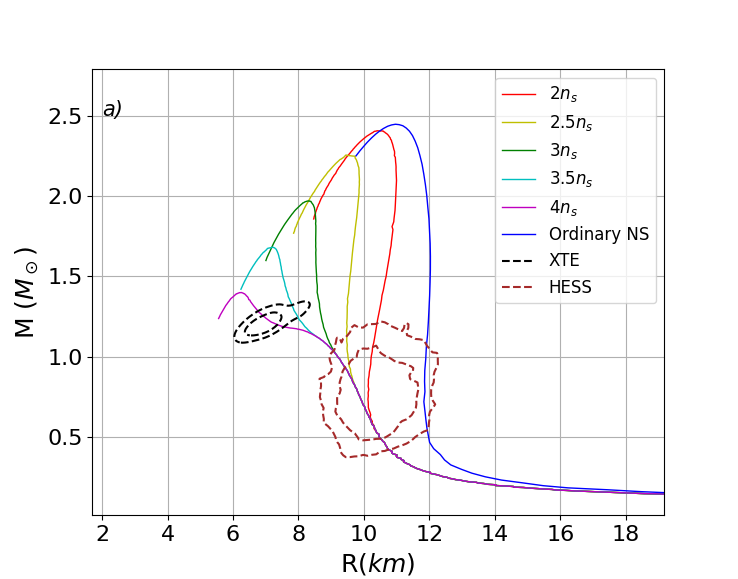}
\includegraphics[width=0.44\textwidth]{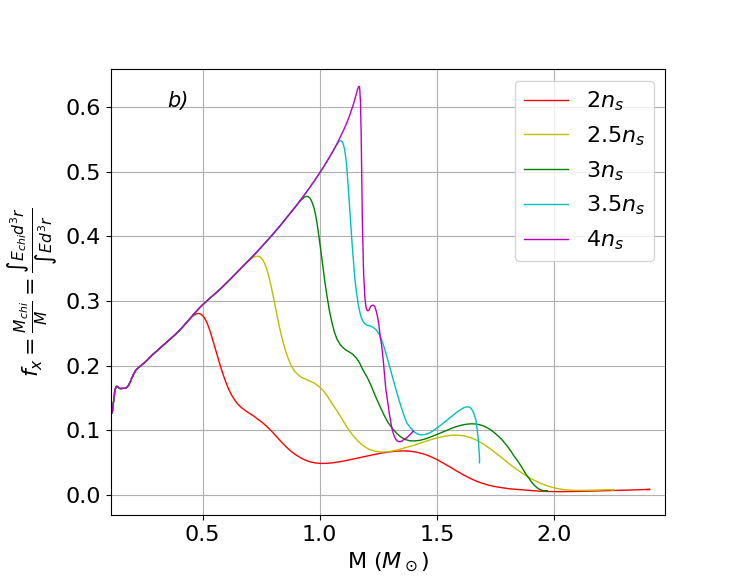}
\caption{a) The M-R dependence and b)  the dark matter mass fraction $f_{
\chi}$  for the case  $z_{\chi}=200$ MeV 
and for various values of the cutoff density, in units of nuclear saturation density $n_s$ where $n_s=0.16$ fm$^{-3}$,   as indicated by the labels. The observational constraints for the HESS J1731-347~\cite{HESS-2023} and for the  XTE J1814-338~\cite{Kini-2024, Baglio-2013}  are also indicated.  }
 \label{zx-ocutof-dens}
\end{figure*}


\begin{figure*}[ht]
\centering
\includegraphics[width=0.44\textwidth]{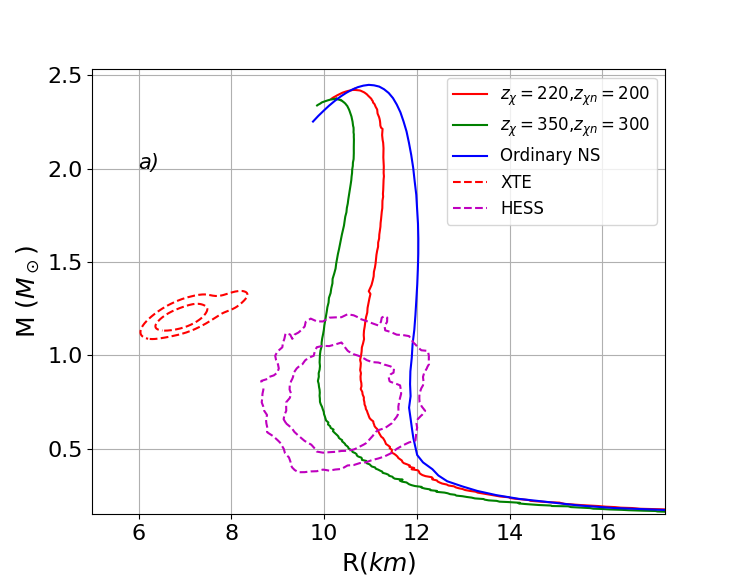}
\includegraphics[width=0.44\textwidth]{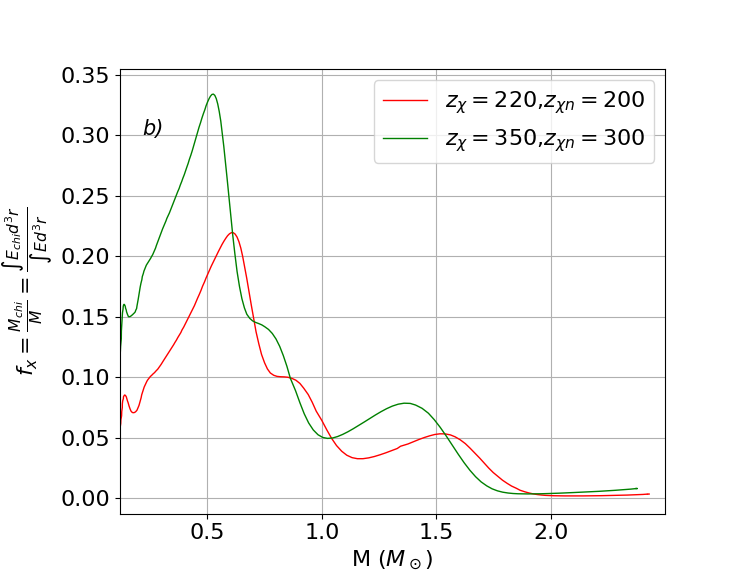}
\caption{a) The M-R dependence and b)  the dark matter  mass fraction $f_{
\chi}$ considering  a cutoff on dark neutron decay for the interaction parameters   $z_{\chi}=220$ MeV and  $z_{\chi n}=200$ MeV (red line)   and  $z_{\chi}=350$ MeV and  $z_{\chi n}=300$ MeV (green line). The cutoff of the dark neutron decay has been specified at twice the nuclear saturation density $n_s$. The observational constraints for the HESS J1731-347~\cite{HESS-2023} and for the  XTE J1814-338~\cite{Kini-2024, Baglio-2013}  are also indicated.  }
 \label{zx-ocutof}
\end{figure*}


\section{Results and Discussion}
In Fig.~\ref{zx-only}, in the left column, we plot the mass–radius dependence for the case with interaction constant 
$z_{\chi}=130$, $z_{\chi}=220$  and $z_{\chi}=1500$ MeV (from the top to the bottom), as well as for the configurations composed of purely neutron matter (ordinary NS) and those involving a mixture of neutron matter and dark matter (compact object). The observational constraints for the HESS J1731-347~\cite{HESS-2023} and for the  XTE J1814-338~\cite{Kini-2024, Baglio-2013}    are also indicated. In the same figures, we also indicate the region in which the number density of dark matter particles becomes dominant (dark matter domination). The key finding is that, with an appropriate choice of the parameter 
$z_{\chi}$, both events can be simultaneously explained as arising from a mixture of dark matter (generated through neutron decay) and neutron matter. Thus, one could potentially argue that the softening of the equation of state, an essential mechanism for explaining these two events, is linked to the dark decay of the neutron.
However, this particular choice of the parameter fails to predict compact objects with masses exceeding two solar masses. It should be clarified here that for very large values of the parameter $z_{\chi}$, the dark matter behaves as an ideal Fermi gas. Since it then constitutes the dominant component at high masses, the maximum mass approaches the well known limit of approximately 0.7 solar masses~\cite{Shapiro:1983du}. 
Furthermore, to elucidate the results, we plotted the dependence of the neutron and dark particle fractions as a function of total density (middle column), as well as the corresponding dark matter mass fraction in each case (right column). In any case, it is evident that at high densities the dark particles dominate, and this has significant implications for the composition of these hybrid compact objects.

Similarly, Fig.~\ref{znx-only} presents the results for the case in which we assume interactions only between neutrons and dark particles. The main finding in this scenario is that the stronger this interaction becomes, the stiffer the equation of state, leading to predictions of higher maximum masses and comparatively large radii.
Furthermore, it is noteworthy that in this case the dark particles outnumber the neutrons in the low density regions, whereas the opposite occurs at high densities.
This is reflected in the mass fraction of dark matter that consists of these objects, which is in all cases smaller than that of the neutrons. This leads to the conclusion that such objects are primarily composed of neutron-rich matter dominating their inner core, with a substantial fraction of dark matter residing in the outer layers of the core. In any case, the key conclusion is that one can predict the existence of compact objects within the mass gap region ($2.5–5\ M_{\odot}$)  whose composition is dominated by neutron-rich matter coexisting with a relatively small fraction of dark matter (see also Ref.~\cite{Vikiaris-2025}). The novel aspect here is that this dark matter is not accumulated over time within the star, but instead originates from it through the dark decay of the neutron.

In Fig.~\ref{zx-zxn} we present the results for the case in which both types of interactions coexist (the self-interaction among dark particles and the interaction between dark particles and neutrons). The results in this scenario are similar to those obtained when only self-interaction is present. A plausible explanation is that the contribution from the dark-matter self-interaction dominates over the interaction between dark particles and neutrons and thus primarily determines the resulting behavior. It should be noted that, with an appropriate combination of the parameters $z_{\chi}$ and $z_{\chi n}$, it is possible to achieve a simultaneous reproduction of both events, although not the observed maximum masses in the vicinity of two solar masses.
It is worth noting here that for very weak interactions (i.e., large values of $z_{\chi n}$), the dark matter component effectively behaves as a free Fermi gas, and the resulting behavior becomes analogous to that shown in Fig.~\ref{zx-only} for large values of $z_{\chi}$.

Additionally, we examine the interesting possibility in which the neutron decay mechanism into dark matter ceases to be operative at higher nuclear matter densities. To the best of our knowledge, this possibility has not been explored in previous related studies, and we consider it to be a particularly interesting scenario that merits further investigation.
In Fig.~\ref{zx-ocutof-dens} we present the results for the case in which the neutron dark decay mechanism ceases, for some reason, to be active. In particular, we present the M–R relation by imposing a cutoff on the dark neutron decay, for the case $z_{\chi}=200,\mathrm{MeV}$ and for several values of the cutoff density.
It is then noteworthy that, depending on the density at which this suppression mechanism sets in, the resulting equation of state becomes relatively soft at low densities and sufficiently stiff at high densities. As a consequence, the predicted mass–radius relation allows for the simultaneous existence of the HESS J1731-347  while also predicting compact objects with masses exceeding two solar masses. It should be noted, however, that the object XTE J1814-338 is difficult to explain within the framework of the mechanism described above. This object is characterized by a relatively small radius, which is challenging to reproduce through any softening mechanism originating from neutron dark decay while simultaneously allowing for the prediction of maximum masses exceeding two solar masses. It is possible that an alternative explanation, in conjunction with existing ones, may shed further light on this issue. Nevertheless, this limitation does not undermine the flexibility of the mechanism we propose. The latter is able to account for the existence of subsolar compact objects over a broad range of masses and radii, while simultaneously preserving the prediction of compact objects with masses equal to or exceeding two solar masses. This addresses the primary shortcoming of previous scenarios.

Finally, in order to further enrich the above analysis, we examined an additional case (see Fig.~\ref{zx-ocutof}), in which, by assuming the cutoff of the dark neutron decay to occur at twice the saturation density, we present the mass–radius relation and the mass fraction $f_{\chi}$ for two different pairs of the parameters $z_{\chi n}$ and $z_{\chi}$. We find that, in both cases, the resulting equation of state becomes soft at low densities and sufficiently stiff at higher densities, leading to a simultaneous prediction of the HESS J1731-
347  constraint and a maximum mass well above the two solar mass limit.
However, one can observe that, in order to also reproduce the object XTE J1814-338, it is necessary to additionally parameterize appropriately the value of the density at which the dark neutron decay is switched off.
In conclusion, by varying the values of the parameters  $z_{\chi n}$, $z_{\chi}$ and the cutoff density,  the model can be made sufficiently flexible with respect to the relevant predictions.

An approach to how two solar mass compact objects could be predicted while simultaneously allowing for the existence of neutron dark decay was proposed in Ref.~\cite{Grinstein-2019}. According to that reference
in case where the neutron decay discrepancies are due to partial neutron decays to DM,  a repulsive interaction between DM and neutrons can disfavor the conversion of neutrons to DM
inside NSs, thus allowing NSs to be heavier than $2 M_{\odot}$, as
supported by observations.
However, under this assumption it is not possible to simultaneously predict compact objects with relatively small masses and radii, whose explanation requires an equation of state that is sufficiently soft at low densities and sufficiently stiff at higher densities.

With regard to the conjecture that neutron dark decay is suppressed at high densities, we list below, without further analysis, few possible mechanisms that may give rise to such behavior.   {\it Additional degree of freedom}: The emergence of additional subatomic degrees of freedom, such as kaons, hyperons, or even quarks, may drastically limit, or even completely suppress, the appearance of dark particles at supranuclear densities. {\it Strong repulsive interaction}:  A possible strong density dependence of the repulsive interaction among dark particles may also play a role. In such a case, the production of dark particles is dramatically suppressed at high densities.  In any case, a systematic investigation of the above mechanisms, or possibly even certain others that may also be taken into consideration, is required to determine not only whether such a suppression of the decay occurs, but also to what extent it takes place.

\section{Concluding Remarks}
The main conclusions of the present study can be summarized as follows: 
a) The proposed mechanism of neutron dark decay, if it exists, may potentially play a significant role in the structure and fundamental properties of neutron stars. In this context, key parameters include, in addition to the mass of the corresponding dark particle, both the strength of the self-interaction among dark particles and the interaction between baryons (neutrons in the present study).
In this case, and assuming a fixed mass for the dark matter particle, we find that exotic compact objects with small masses and radii can be explained through an appropriate choice of the parameter governing the repulsive self-interaction. However, this requires the assumption that the existence of such objects is accounted for within the framework of the two branch scenario,  that is, the simultaneous prediction of both a high maximum mass and the presence of exotic compact objects is not possible unless one resorts to the two-branch interpretation (see also the relevant  Refs.~\cite{Veselsky-2025b,Brown-Rho_2008,Haensel_2007}). When only the interaction between dark particles and neutrons is taken into account, and, in the particular case where this repulsive interaction becomes very strong (small values of $z_{\chi n}$), we find configurations that include objects with significantly large maximum masses and correspondingly large radii (compared to conventional neutron star models), as a consequence of the emergence of very stiff equations of state. It is noteworthy that, in this case, the dark matter fraction is substantially smaller than the neutron mass fraction. If both types of interactions are taken into account simultaneously, the results are similar to those of the first case. In this scenario, however, the additional freedom associated with the two interaction parameters renders the resulting configurations more complex and flexible. b) An additional important factor, considered for the first time in the present study, concerns whether the decay mechanism is active throughout the entire core of the star. This assumption is motivated by the conjecture that, although this mechanism has been proposed in the context of terrestrial laboratory conditions (i.e., low densities), it remains unclear whether, and to what extent, it operates at the much higher densities prevailing in the interior of neutron stars. This conjecture is therefore explicitly taken into account in the present analysis. In this case, we obtain a particularly interesting result by also taking into account the density range over which the decay mechanism is active. Specifically, it is possible to generate configurations in which neutron dark decay coexists with a consistent prediction of both the two–solar-mass limit and the presence of compact objects with subsolar masses, without resorting to the assumption of two distinct branches.



\end{document}